\documentclass[12pt,letterpaper]{article}

\usepackage[numbers,sort&compress]{natbib}

\usepackage{datetime}
\usepackage{amsmath,amssymb,array,calc,rotating,epsfig,psfrag, amscd}
\usepackage{color}
\usepackage[colorlinks=false,
    linktoc=all, 
    pdfstartview=FitV,
    bookmarksopen=true]{hyperref}
\usepackage[left=2cm,top=1cm,right=3cm,nohead]{geometry}
\usepackage[english]{babel}

\newcommand{\bea}{\begin{eqnarray}}
\newcommand{\eea}{\end{eqnarray}}
\newcommand{\be}{\begin{equation}}
\newcommand{\ee}{\end{equation}}
\newcommand{\barr}{\begin{array}}
\newcommand{\earr}{\end{array}}

\newcommand{\tphi}{\tilde{\phi}}
\newcommand{\txi}{\tilde{\xi}}

\newcommand{\csch}{{\rm csch \,}}

\newcommand{\non}{\nonumber}
\definecolor{cardinal}{rgb}{0.6,0,0}
\definecolor{darkgreen}{rgb}{0,0.5,0}
\definecolor{golden}{rgb}{0.92, 0.7, 0}
\definecolor{midnight}{rgb}{0, 0, 0.5}
\definecolor{darkblue}{rgb}{0.2, 0, 0.8}

\newcommand{\beq}{\begin{equation}\begin{aligned}}
\newcommand{\eeq}{\end{aligned}\end{equation}}

\def\IS{\ensuremath{\mathbb S}}

\def\IZ{{\mathbb Z}}

\numberwithin{equation}{section}

\begin{document}

\thispagestyle{empty}
 \begin{flushright}
  IPhT-t13/222
 \end{flushright}
\begin{center}
\end{center}
\vspace{0.5cm}
\begin{center}
\baselineskip=13pt {\LARGE \bf{Uplifting the baryonic branch: a test \vskip0.1cm  for   backreacting anti-D3-branes}}
 \vskip1.5cm 
\renewcommand{\thefootnote}{\fnsymbol{footnote}}
Anatoly Dymarsky$^a$  and Stefano Massai$^{b,c}$\\
\vskip0.5cm
 \textit{$^a$ DAMTP, University of Cambridge,\\ 
Cambridge, CB3 0WA, UK}\\
\href{mailto:dymarsky@damtp.cam.ac.uk}{dymarsky@damtp.cam.ac.uk}
\vskip0.5cm
 \textit{$^b$ Institut de Physique Th\'eorique,\\
  CEA Saclay, F-91191 Gif-sur-Yvette, France}\\
\vskip0.5cm
\textit{$^c$ Arnold Sommerfeld Center for Theoretical Physics,\\
  Theresienstr. 37, 80333 Muenchen, Germany}\\
\href{mailto:stefano.massai@lmu.de}{stefano.massai@lmu.de}
\vskip0.5cm
\end{center}

\begin{abstract}
\noindent
Placing D3 or anti-D3-branes at the tip of 
the Klebanov-Strassler background results in uplifting the
baryonic branch of the moduli space of the dual field theory. In this paper we derive a mass formula for the scalar particle associated with the motion along the baryonic branch,  from both open and closed string points of view. We show that both methods give the same mass at linear order in number of (anti)D3-branes, thus providing a comprehensive check for the recently found linearized supergravity solution describing  backreacting anti-D3-branes at the tip. 

\end{abstract}

\clearpage


\section{Introduction}\label{sec:intro}

For the last thirteen years Klebanov-Strassler (KS) background \cite{Klebanov:2000hb} enjoys the role of a default playground to test various ideas of gauge/gravity duality, whenever maximally supersymmetric ${\mathcal N=4}$ YM theory wouldn't suffice. In particular in 2001 Kachru, Pearson and Verlinde suggested that placing anti-D3-branes  at the tip of the KS geometry would lead, through the  backreaction of anti-D3s, to a holographic background dual to a SUSY-breaking metastable state in the dual field theory \cite{KPV}. This so-called KPV metastable state is peculiar from the field theory point of view. Conceivably because of the absence of a small dimensionless parameter characterizing this state, so far it was  out of reach for all known field theoretic techniques. At the same time the corresponding hypothetical background is a key ingredient in the most explicit scenario proposed to date that achieves a four-dimensional de-Sitter in a String Theory compactification \cite{KKLT}. Since the holographic background in question was never constructed explicitly, presumably because of its complexity, the fate of the  String Theory  landscape of the $dS$ compactifications is currently resting on the original probe D-brane calculation \cite{KPV}. In light of the importance of the question it is only natural to wish to obtain the backreacted supergravity background explicitly. This ambitious task was first seriously examined in \cite{Bena:2009xk} where the authors put forward all necessary ingredients to fully investigate $SU(2)\times SU(2)\times \mathbb{Z}_2$ invariant sector of the background in question at the linear order in number of anti-D3-branes (the earlier attempts \cite{Chen:2009zi, DeWolfe:2008zy} used various approximations which hamper their ability to determine the desired background simultaneously near and far away from the anti-D3-branes). The original work  \cite{Bena:2009xk} revealed complexity of the problem arising already at the linearized level, including subtlety of formulating boundary conditions that would correspond to the desired KPV state. This work continued in \cite{Dymarsky:2011pm,Bena:2011wh,Bena:2011hz} until the $SU(2)\times SU(2)$ invariant mode of the corresponding linearized solution was obtained in the most possible explicit form.

Now, once the linearized solution is found, it is very desirable to submit it to all possible self-consistency tests and checks, especially because many technical details associated with formulating boundary conditions etc.~were at the time a subject of a heated debate. A simplest consistency check comes from studying interaction potential between a number of anti-D3-branes sitting at the tip of the KS geometry and a stack of D3-branes located some finite distance away. At the linearized order in number of both D3s and anti-D3s there are two ways to proceed. One is to treat D3s as sources backreacting on the geometry while anti-D3s would be probes \cite{KKLMMT}. Another way is to treat D3s as probes in the (linearized) geometry created by the backreacting anti-D3s. Both calculations yield the same result \cite{Bena:2010ze,Dymarsky:2011pm}, but unfortunately this test is only sensitive to the IR boundary condition for just one (out of many) supergravity mode. Hence this hardly passes as a very comprehensive check.  

Another check of the linearized solution for anti-D3-branes comes from comparing vacuum energy of the dual state in field theory (the ADM mass in supergravity) with the corresponding probe calculation. This check is sensitive to both IR and UV behavior of the solution but certainly does not involve every single part of it; it is possible to show that the solution in question passes this check, without actually obtaining all parts of the solution explicitly \cite{Dymarsky:2011pm}. In this paper we propose and perform yet another check which  is sensitive to all parts of the solution in question. 

The idea of our check is to study the effect of anti-D3-branes on the so-called baryonic branch of moduli space from both open and closed string points of view. Before we go into details, let us take a quick detour and remind the reader basic facts about the KS solution which are relevant to what follows. The field theory dual to the KS   background is a particular ${\mathcal N}=1$ $SU(N+M)\times SU(N)$ SYM with $N$ divisible by $M$. This theory admits a so-called baryonic branch of moduli space and in fact the KS background is dual to a particular $\IZ_2$ symmetric vacuum located at the locus  of the baryonic branch. There is a whole one-dimensional family of distinctive vacua in field theory and there is a family of supergravity solutions dual to them in a holographic sense \cite{Aharony:2000pp,GHK,BGMPZ}. The KS background is only a particular solution from this family. Adding D3 or anti-D3-branes to the KS geometry changes $N$ and as a result the baryonic branch gets uplifted i.e.~the massless scalar particle associated with the motion along the branch  becomes massive.  The corresponding emergent potential near the origin of the branch can be calculated with ease up to linear order in number of (anti)D3-branes using probe  approximation \cite{DKS}. The idea of this paper is to perform the same calculation using closed string channel i.e.~by taking the backreacted solution describing (anti)D3-branes and studying the mass of the deformation along the direction in the field space associated with the motion along the baryonic branch. As advertised in the abstract the linearized solution describing the KPV state passes this test with the flying colors.

Before we proceed with the technicalities let us point out that the machinery developed in the paper to study the uplift of the baryonic branch due to (anti)D3-branes is universal in a sense that it can be used to study the uplift due to any $\IZ_2$-invariant small perturbation of the KS background.\footnote{This is more general than small perturbations (by various operators) of the dual theory as here we also assume one can add small number of D-branes which, strictly speaking, change the dual field theory in a non-continuous way.} To better explain the logic of our calculation we in fact start by considering the uplift of the baryonic branch associated with perturbing the original ${\mathcal N}=1$ theory by SUSY-breaking gaugino masses. 

This paper is organized as follows. In section 2 we develop the machinery necessary to calculate the uplift of the baryonic branch at the origin (mass of the scalar associated with motion along the branch) for any $SU(2)\times SU(2)\times \mathbb{Z}_2$ perturbation of the KS background. In section 3 we apply these results to calculate the uplift due to infinitesimally small gaugino masses. In section 4 we calculate the uplift for (anti)D3-branes placed at the tip from both open and closed string points of view and compare the results thus testing the linearized solution describing the KPV state. We discuss our findings in section 5. Technical details are delegated to the appendices.

\section{Baryonic branch uplift}\label{sec:bbuplift}
In this section we develop the machinery necessary to calculate the uplift of the baryonic branch near the origin, due to arbitrary $\IZ_2$-even perturbation.
\subsection{Baryonic branch from supergravity}

As we briefly mentioned in the introduction the KS solution is only one particular point from the so-called BGMPZ family of supergravity solutions \cite{BGMPZ} which share the same leading UV-asymptotic and hence correspond to different vacua of the same filed theory. The vev $U=\langle \hat{U}\rangle$ of the bottom component of the baryonic $U(1)$ current  
\bea
\hat{U}= Tr(|A|^2-|B|^2)\ ,
\eea  is usually chosen to be the parameter along the branch. The point $U=0$ corresponds to the KS solution. There is a $\IZ_2$ symmetry acting as $U\rightarrow -U$ leaving the KS solution invariant (the geometric meaning of this symmetry is discussed in \cite{KW}). From the supergravity point of view
infinitesimal motion along the branch near the origin $U=0$ is associated with  the $\IZ_2$-odd linear deformation of the KS solution that satisfies certain boundary conditions in the UV. The corresponding  supergravity mode $\tilde{z}(\tau)$ was first found by Gubser, Herzog and Klebanov (GHK) in  \cite{GHK}. It satisfies a Schr\"odinger-type equation in the radial direction of the conifold $\tau$
\bea
\label{Sh}
-\tilde{z}''+V_{\tilde{z}}(\tau)\, \tilde{z}=0\ .
\eea    
Zero eigenvalue in \eqref{Sh} is due to the fact that motion along the baryonic branch costs no energy i.e.~the branch is flat. 

The particular form of $V_{\tilde{z}}$ in \eqref{Sh} (see section~\ref{upliftmachinery}) leads to the following general solution for $\tilde{z}$ in the UV region $\tau\rightarrow \infty$ 
\bea
\label{bc}
\tilde{z}=\alpha (\tau-\tau_0)+\beta\ ,
\eea
where $\alpha$ and $\beta$ are some constants, and the constant $\tau_0$ is not physical as it always can be absorbed into $\beta$. Taking into account that $\tau \sim \log r$ where $r$ is the ``AdS'' radius of the conifold we find that \eqref{bc} is consistent with the dimension $2$ of the operator $\hat{U}=Tr(|A|^2-|B|^2)$. Usually $\alpha$ would correspond to the coupling and $\beta$ to the vev of $\hat{U}$. But in this particular case these roles happen to be reversed: $\alpha$ is the vev and $\beta$ is the coupling \cite{GHK}. Although this is somewhat unusual there is no contradiction with the AdS/CFT dictionary because for the dimension $2$ operator both interpretations are valid \cite{KW2}. The asymptotic of the GHK mode associated with the motion along the branch is $\tilde{z}\sim (\tau-1)$ which means the vev $\alpha\sim U$ changes while the coupling $\beta$ remains zero. 

\subsection{Boundary conditions: a puzzle and a resolution}
\label{bcsection}
What happens if we deform the KS background by introducing some small $\IZ_2$-even perturbation? Unless we are lucky this will result in the baryonic branch uplift i.e.~a non-trivial potential $V(U)$ emerging along the branch
\begin{equation}
\label{puplift}
V(U) = V(0)+m^2_U U^2 + {\mathcal O}(U^4)\, .
\end{equation} 
The mass $m^2_U$ will appear as an eigenvalue in the equation \eqref{Sh} as a response to a new effective potential $V_{\tilde{z}}(\tau)$. Below we will develop a machinery to calculate new $V_{\tilde{z}}$ for any $\IZ_2$-even perturbation of KS and hence the problem of calculating $m^2_U$ will reduce to some trivial numerics. 

Our logic is clear  and seems to be completely infallible but in fact we will immediately run into a puzzle if we consider D3-branes as a perturbation. One can be absolutely certain that the D3-branes sitting anywhere on the conifold  create a potential of the form \eqref{puplift} with some non-zero $m^2_U$ \cite{DKS}. At the same time direct calculation or reasoning based on the particulars of the GHK geometry\footnote{The deformed conifold,  the geometry behind the KS solution, is Ricci-flat. In fact the metric of the GHK perturbation is also Ricci-flat, as can be deduced from the constant dilaton of the GHK solution (also see \cite{Dymarsky:2011ve} for a detailed discussion). For such a background the supergravity equations factorize \cite{GKP}: the only remaining equation is one for the warp-factor $h$. Presence of D3-branes affects the warp-factor  through the equation $\nabla^2 \delta h=\sum \delta_{D3}$. At the linear  order in $U$ the $\IZ_2$-odd GHK perturbation of the unwarped metric can not affect $\nabla^2$ because  it is $\IZ_2$-even. Hence the GHK background with the new warp-factor solves the supergravity equations of motion in the presence of D3-branes up to quadratic order in $U$. Alternatively, one can see that presence of D3-branes does not affect $V_{\tilde{z}}$ in \eqref{Sh} because the only part of the background affected by the D3-branes is the warp-factor, but the explicit expression for $V_{\tilde{z}}$ (it is derived later in the text) is warp-factor independent.} reveal that $V_{\tilde{z}}$ does not change and hence the same GHK mode still solves the equation \eqref{Sh} with zero eigenvalue. The only possible interpretation of this would be that D3-branes do not uplift the branch, at least at the lowest order  in $U$. This puzzling contradiction is something we have to resolve before turning to a more complicated case of anti-D3-branes. 

It is tempting to attribute the mismatch between the probe and SUGRA calculations to a singular nature of the supergravity background whenever D3-brane are present. This would certainly remove the puzzle but a detailed consideration shows that the suspected singularity of $\delta h$ is not invalidating supergravity-based  calculation. Instead of providing a detailed justification of this fact, let us instead 
sharpen the puzzle by considering another confusing example when the supergravity background remains smooth and weakly curved. Let us perturb the dual gauge theory by the infinitesimally small gaugino mass $m\lambda^2$ by giving a non-zero value to the top component of the gauge coupling superfield $\tau={i\over g^2}+\dots + \theta^2 m$
\bea
 {\mathcal L}=\int d^2 \theta\, \tau\, W_{\alpha}^2\ .
\eea Here $W_{\alpha}$ is a vector superfield. Presence of the top component in $\tau$ breaks supersymmetry, but at the linear order  in $m$ holomorphy implies that the usual expression for the low-energy superpotential still holds $W\sim e^{i\tau}$. The resulting potential $V=\int d^2\theta\, W$  acquires the following linear-in-$m$ contribution
\bea
\label{dV}
\delta V=m \Lambda^3+{\rm c.c.}
\eea
The internal scale of the field theory $\Lambda^3=\langle \lambda^2\rangle$ does not depend on the choice of vacuum along the baryonic branch \cite{DKS} and therefore \eqref{dV} contributes to $V(0)$ in \eqref{puplift}, but not to $m^2_U$ which remains zero at linear order in $m$. One can arrive at the same conclusion by noticing that $m$ has $U(1)_R$ charge $-3$ while $m^2_U$ is $U(1)_R$ invariant. Hence the only possible dependence on $m$ at linear level could be $m^2_U\sim m \Lambda^2/\bar{\Lambda}+{\rm c.c.}$ which is not allowed because of  $\bar{\Lambda}$ in the denominator. 

In the discussion above we were sloppy and disregarded the fact that the dual  gauge theory has not one but two gauge groups and correspondingly two pairs of gauginos. But exactly the same logic shows that $m^2_U$ can not depend on neither of two gaugino masses $m_i$, $i=1,2$. 

What do we see on the supergravity side? Two linearized  supergravity backgrounds dual to the KS theory perturbed by the infinitesimal gaugino masses were explicitly found in \cite{Kuperstein:2003yt, Dymarsky:2011pm, Bena:2009xk}. If we calculate new $\delta V_z$ we will see that it is non-zero and correspondingly there is no reason to believe new \eqref{Sh} admits zero eigenvalue solution any more. The new eigenvalue at linear order in $m_i$ is thus
\bea
\label{gmu}
m^2_U=a_i m_i\ ,
\eea  
where $a_i$ are some numerical coefficients. One can clearly combine $m_1, m_2$ such that \eqref{gmu} vanishes but in a general case the gaugino masses induce a non-vanishing uplift of the baryonic branch. Thus we again run into a contradiction: the supergravity calculation does not match the dual field theory analysis. 

We believe by this point we have intrigued our reader enough. The situation seems to be very puzzling: in one case the field theory suggests the uplift should be zero, but supergravity produces a non-vanishing answer. In the other case  probe calculation gives a non-zero result, but supergravity  suggests the uplift is not there. 

The resolution of the puzzle comes from a careful treatment of the boundary conditions satisfied by the ``wave-function'' $\tilde z$ in \eqref{Sh}. At first glance this sounds very surprising. Indeed, in case of D3-branes we argued that exactly the same solution that solves \eqref{Sh} without D3-branes will still solve \eqref{Sh} after D3-branes are added. Obviously the same solution has the same boundary conditions. This makes a lot of sense: the boundary condition for $\tilde z$ should always be the same -- coupling $\beta$ in  
\eqref{bc} should remain zero while vev $\alpha$ can vary. The subtlety we are missing is due to $\tau_0$ in \eqref{bc} which can change  what we call $\beta$. In AdS/CFT the boundary conditions are to be formulated at a holographic screen located at some large physical radius $r_{\rm cutoff}$ which morally corresponds to some physical UV cut-off in field theory. The radial variable $\tau$ at the same time is dimensionless, related to $r$ though the conifold deformation parameter $\varepsilon$
\bea
\label{taur}
\tau = \log( r^3 /\varepsilon^2)+O(r^{-1})\ .
\eea 
Since $\varepsilon$ is the only dimensionful parameter of the solution,  no dimensionless quantity can depend on it and accordingly $\varepsilon$ is commonly taken to be one. But this situation changes if one considered two holographic backgrounds, the original and the perturbed one. Say, we start with the KS background and put the holographic screen at $r_{\rm cutoff}$. Based on the behavior of $\tilde z$ at large $r$
\bea
z=\alpha (\log(r^3/\varepsilon^2)-1)+\beta\ ,
\eea
and comparing with the known solutions describing baryonic branch \cite{GHK,BGMPZ} we conclude that 
\bea
\label{alpha}
a:=\left.{dz\over d\log r^3}\right|_{r=r_{\rm cutoff}}\ ,
\eea
is a vev, while
\bea
\label{beta}
b:=\left. z(r)-a  (\log(r^3/\varepsilon^2)-1) \right|_{r=r_{\rm cutoff}}\ ,
\eea
is a coupling of $\hat{U}$. Now, if we consider a slightly perturbed background with new $\varepsilon$ we still want to keep the {\it same boundary conditions} at the {\it same radius} $r_{\rm cutoff}$, namely \eqref{beta} to stay zero, while \eqref{alpha} can be arbitrary. As one can see the effect of changing $\varepsilon$ will be in shifting the definition of $\beta$.

Let us now return to the puzzling examples discussed above. The pure KS background and the KS background with $p$ D3-branes correspond holographically to different gauge theories which have completely different (in fact totally unrelated) $\varepsilon$. But if $p=\ell M$ for some integer $\ell$ these two backgrounds describe two different vacua of the same theory in which case the new and the old $\varepsilon$'s are related by \cite{DKS,Dymarsky:2011pm}
\bea
\label{phys}
\varepsilon_{M}^2=\varepsilon^2\, e^{2\pi \ell\over M}\ .
\eea
Based on this relation we {\it conjecture} that choosing 
\bea
\label{conjecture}
\varepsilon_p^2 =\varepsilon^2\, e^{2\pi p\over M^2}\ .
\eea
in presence of any number $p$ of D3-branes will define correct boundary conditions for $\tilde{z}$ by identifying \eqref{alpha} with the vev and \eqref{beta} with the coupling of $\hat{U}$ (these formulae implicitly define  $\tau_0$). Let us emphasize: while \eqref{phys} is a relation of deformation parameters for two distinct solutions (which correspond to different vacua fo the same field theory) and is proven to be correct,  \eqref{conjecture}, together with \eqref{alpha} and \eqref{beta} is a conjecture about holographic dictionary for the $\tilde{z}$ mode.  
The test of the backreacted anti-D3-branes solution advertised in the introduction should be understood as a check of this conjecture as well. 

It becomes clear now why supergravity approximation did not ``see'' the mass $m^2_U$ in case D3-branes were added. Indeed the same GHK mode  with the same $\alpha, \beta, \tau_0$ solves the supergravity equations of motion. But since $\varepsilon$ is different, for $\alpha,\beta$ to have the same interpretation in terms of vev and coupling, $\tau_0$ should now change. Hence the same GHK solution has different interpretation now: it corresponds to the mode with some non-trivial vev $U$ and some non-zero coupling of $\hat U$. This solution describes not only D3-branes but also perturbation of theory by $\hat{U}$! In fact there is another solution of \eqref{Sh} with the non-zero eigenvalue $m^2_U$ and vanishing \eqref{beta}. This is the correct solution  describing motion along the baryonic branch when the only new ingredient is a stack of D3-branes.  

The supergravity solutions describing infinitesimal perturbation of KS by gaugino masses also have new value of $\varepsilon$. As a result to find corresponding eigenvalue $m^2_U$ one has to solve \eqref{Sh} with the new potential $V_z$ and the new boundary conditions, requiring  \eqref{beta} to be zero. As we will show later a small miracle happens and the effect of the new potential cancels against the effect of the new boundary conditions such that there is a zero mode solution for any $m_i$, confirming that $m^2_U$ remains zero, as expected.  

\subsection{The baryonic branch uplift from supergravity}
\label{upliftmachinery}
Now we are ready to calculate the appropriate
equation for the ``wave-function'' $\tilde{z}$ of the $U$ mode -- the generalization of \eqref{Sh} for the more general backgrounds. 
First, we notice there is only one $\IZ_2$-odd linearized mode which has the appropriate UV asymptotic corresponding to dimension 2 and correspondingly we have to find the equations describing fluctuation of this mode (and all other modes it couples to) in some abstract $\IZ_2$-even background. Such equations were
derived in~\cite{Benna:2007mb} for the Klebanov-Strassler background,
with the purpose of studying the mass spectrum of certain scalar
glueballs. Here we will generalize this result to a full $\IZ_2$-even
$SU(2)\times SU(2)$ symmetric ansatz (colloquially called the PT-ansatz after Papadopoulos and Tseytlin \cite{PT}). The only equation we need here is (the details can be found in Appendix~\ref{appVzder})\footnote{A similar result was previously obtained by C.~Ahn and T.~Tesileanu.}:
\begin{equation}\label{zfulleom}
\tilde z'' - V_{\tilde z}  \tilde z = -m^2 \Big[e^{-2(A+4p)} \tilde z+\frac{3\sqrt{6}
    P}{4}e^{-A-4p-x-\frac{\Phi}{2}}(f' e^{-y} +k' e^{y}) \tilde \omega \Big]\, ,
\end{equation}
where $V_{\tilde z}$ is the following function of the PT scalars (in Einstein frame):
\begin{align}
V_{\tilde z} &=\frac12 \cosh (2y) +\frac12 e^{-4x} \Big[2 e^{-12p}-e^{4x}+e^{2x-2y+\Phi}(e^{2y}(2P-F)+F)^2\Big] \non\\
& \quad -\frac12 e^{-2(x+y)-\Phi}(f'+e^{2y}k')^2-4(A'+p')^2+y'^2 \, .\label{vz}
\end{align}
Here $m^2$ is the mass of the baryonic branch parameter $U$, it's different from $m^2_U$ only by some normalization factor. 

Above we have introduced  
\begin{equation}
\tilde z = 4\, z\, e^{2A + 2p} \, , \qquad \tilde \omega= \frac{\sqrt{2}}{3 P\sqrt{3}} e^{A+10p+3x-\Phi/2} \, ,
\end{equation}
using the original GHK wave-functions $z$ and $\sigma'$ from~\cite{GHK,Benna:2007mb}.
When restricted to the KS solution with $\varepsilon$ taken to be one, $V_{\tilde z}$ reduces to
\begin{equation}
V_{\tilde z} ^{KS} = 2\sinh^2 y_0= \frac{2}{\sinh^2\tau} \, ,
\end{equation}
and the massless equation is solved by the GHK mode
\begin{equation}\label{solz0}
\tilde z_0 = \tau \coth\tau-1 \, .
\end{equation}
To compute the mass for a general linear perturbation around the
Klebanov-Strassler background, we expand the potential $V_z$ as
\begin{equation}
V_{\tilde z}  = V_{\tilde z} ^{0} + \delta
V_{\tilde z}  \, ,
\end{equation}
and take $V_{\tilde z} ^0=V_{\tilde z}^{KS}$. We use the $0$ subscript for the modes of the unperturbed solution.
Since the mass of the $U$ mode vanishes at the zeroth-order, we don't need to compute linear fluctuations of the mode $\tilde \omega$, and as a consequence the equation for $\tilde z$ decouples.

We begin by multiplying both sides of~\eqref{zfulleom} by $\tilde z_0$ and 
integrate from $0$ to
$\infty$:
\begin{equation}\label{integratezeom}
 \int_0^{\infty} \Big[\tilde z_0 \tilde z'' - V_{\tilde z}^0 \tilde z_0 \tilde z -\delta V_{\tilde z} \tilde z_0^2 \Big]d\tau= -m^2 \kappa_0
 \, ,
 \end{equation}
where $\kappa_0$ is the coefficient of $m^2$ in~\eqref{zfulleom} evaluated at the zeroth-order:
\begin{equation}
\kappa_0 =  \int_0^{\infty} \Big[e^{-2(A_0+4p_0)} \tilde z_0^2+\frac{3\sqrt{6}
    P}{4}e^{-A_0-4p_0-x_0}(f_0' e^{-y_0} +k_0'
  e^{y_0}) \tilde z_0\tilde \omega_0 \Big]d\tau \, .
\end{equation}
This constant can be evaluated numerically (see below). 
 We
observe that from~\eqref{zfulleom} we have at the zeroth-order
\begin{equation}
\frac{2}{\sinh^2 \tau} \tilde z_0 = \tilde z''_0 \, ,
\end{equation}
so we can integrate by parts the first two terms in the left-hand-side, to obtain
\begin{equation}
\int_0^{\infty} \Big[\tilde z_0 \tilde z'' -\frac{2}{\sinh^2 \tau}
  \tilde z_0 \tilde z\Big] d\tau = \int_0^{\infty} \Big[\tilde z_0 \tilde z'' -
  \tilde z''_0 \tilde z\Big] d\tau = \Big[ \tilde z_0 \tilde z' -\tilde z'_0 \tilde z
  \Big]_{0}^{\infty} \, .
\end{equation}
Note that the UV asymptotic of the original
wave-function~\eqref{solz0} is $\tilde z_0 \sim \tau -1$, while for the
perturbed solution $\tilde z$ we allow for a general behavior
\begin{equation}
\tilde z \sim \tau + \delta \tau -1 \, .
\end{equation}
Here
the shift in $\tau$ coordinate $\delta \tau$ is the same as new $\tau_0=1-\delta\tau$. As we explained in section \ref{bcsection} it is related to the shift in the deformation
 parameter of the conifold $\varepsilon$ for the given background in question.
Noticing that
the IR limit does not contribute we find
\begin{equation}
\Big[ \tilde z_0 \tilde z' -\tilde z'_0 \tilde z
  \Big]_{0}^{\infty} = -\delta \tau \, .
\end{equation}
The mass formula we obtain is thus:
\begin{equation}\label{massfinalexprs}
\delta \tau + \int_0^{\infty} \delta V_{\tilde z} \, \tilde z_0^2\,  d\tau= m^2 \kappa_0
\, .
\end{equation}

We now discuss how to compute various quantities that enter 
the formula above. First, we would like to determine the shift $\delta \tau$, which is done by matching running of D5-charge in the UV (compare with \eqref{bc})
\bea
Q^{D5}(r)={\sf a}\tau+{\sf b}\ .
\eea
The idea behind the calculation of $\delta \tau$ is the following. One requires that 
two solutions which correspond to different states in the same theory have the same charge $Q^{D5}$ at some large radius $r_{\rm cutoff}$\footnote{Strictly speaking the KPV state described by backreaction of $p$ anti-D3-branes is a state in $SU(N+M+(M-p))\times SU(N+(M-p))$ theory. Therefore to calculate $\delta \tau$ for that solution we would need to guess which $\delta \tau$ is caused by adding $M-p$ D3-branes to the KS background. This is exactly what we did in \eqref{conjecture}}. Through $\tau=\log(r^3/\varepsilon^{2})$ change of $\varepsilon$ introduces the shift $\tau\rightarrow \tau+\delta\tau$ which in turn should be compensated by the change of $\sf b$~\cite{Dymarsky:2011pm}. 
In the notations introduced in \cite{Borokhov:2002fm,Bena:2009xk}  linear perturbation of the  D5-charge at infinity is controlled by only one mode  $\tphi_5$ which approaches a constant at infinity
\begin{equation}\label{QD5UV}
Q^{D5}(\tau) \sim P(\tau -1) + \tphi_5(\infty) \, .
\end{equation} 
Hence equivalence of $Q^{D5}(r_{\rm cutoff})$ for two  backgrounds  (the original and the perturbed one) takes the form
\begin{equation}
P (\tau_0 -1) = P (\tau_0 + \delta \tau -1 ) + \tphi_5(\infty) \, .
\end{equation}
From here we get
\begin{equation}\label{deltatauphi5}
\delta \tau = -\frac{\tphi_5(\infty)}{P} \, .
\end{equation}
So the shift in $\tau$ can be unambiguously computed for the 
solutions of interest, such as perturbation by the gaugino masses, or backreaction of anti-D3-branes.

Second, we obtain the expression for $\kappa_0$ numerically
\begin{equation}\label{dI}
\kappa_0 = \int_0^{\infty} \Big[\frac{h_0}{6}\frac{\sinh^2 \tau}{(\cosh \tau
  \sinh \tau -\tau)^{2/3}} \tilde z_0^2 -\frac{6P^2}{\sinh \tau} (\cosh \tau
\sinh\tau-\tau)^{1/3}\tilde z_0 \tilde \omega_0 \Big] \approx 171.583 P^2 \, .
\end{equation}

Third, by expanding \eqref{vz} up to linear order we find that $\delta V_{\tilde z}$
can be expressed in terms of the perturbation modes $\tilde \xi^a$ and $\tilde \phi^a$ introduced in~\cite{Borokhov:2002fm} as follows:
\begin{align}\label{Vzxiphi}
\delta V_{\tilde z} & = 2\sinh (2y_0)\, \tilde \phi_2
+ e^{-4(A_0+p_0)}\Bigg[4 \sinh y_0 \, \txi_2 + \frac23 e^{-6p_0 -2x_0}(2\tilde \xi_1 +\tilde \xi_3 +\tilde \xi_4) \non\\
&\quad -8\cosh y_0 (e^{y_0} P - F_0 \sinh y_0) \txi_5 - 8\sinh y_0 (e^{y_0} P - F_0 \sinh y_0) \txi_6 \Bigg]\, .
\end{align}
This form  is instrumental since we can now use the explicit numerical solution of~\cite{Bena:2011wh} to evaluate the above expression.
Given a particular perturbation of KS,
formulae~\eqref{deltatauphi5},~\eqref{dI}, and~\eqref{Vzxiphi}  allow us 
to compute the baryonic branch uplift (the mass of the particle associated with the motion along the branch). In the next section we will perform such a  calculation for various solutions of interest.

\section{Gaugino mass perturbation}\label{sec:kuso}

In this section we discuss the Klebanov-Strassler field theory perturbed by infinitesimally small gaugino masses which softly break
supersymmetry. As was mentioned before there are two pairs of gauginos and hence two masses $m_1, m_2$. Therefore on gravity side we expect two linearized solutions. One gravity solution of this kind was obtained
in~\cite{Kuperstein:2003yt}. It corresponds to some (unspecified) linear combination of $m_1, m_2$. The second linearly independent solution  was obtained in~\cite{Dymarsky:2011pm,Bena:2011wh}.

In what follows we use the notations of \cite{Bena:2011wh} for both solutions. Thus the two independent parameters of the linearized gravity solutions are denoted by $X_2$ and $X_7$ that are related to $m_1, m_2$ by an unspecified linear transformation.
The supersymmetry breaking modes $\tilde \xi^a$ of the 
solutions in question are known analytically 
\allowdisplaybreaks{
\begin{align}
\txi_1 & = \txi_3 = \txi_4 = 0 \, , \label{xi2gauginomasses} \\
\txi_2 & = \frac{X_2}{4}\csch^3 (\tau) (\sinh (2\tau ) - 2 \tau)^2
\non \\
& \qquad - P
X_7 \Big[\csch (\tau) + \tau \big(\cosh(\tau) - 2 \coth(\tau) \csch(\tau)
+ \tau \csch^3 (\tau)\big)\Big] \, ,\non\\
\txi_5 & = \frac{X_2}{P}-\frac{3X_7}{2} \, , \non\\
\txi_6 & = \frac{X_2}{P} \tau \csch(\tau) + X_7
\left[-\frac12 \cosh(\tau) - \tau \csch (\tau)\right] \, , \non\\
\txi_7 &=  \frac{X_2}{P}\Big[  \tau \coth(\tau) -1\Big] \csch (\tau)
\non \\
& \quad + X_7
\left[\csch(\tau) - \tau\coth(\tau)\csch(\tau) + \frac12
  \sinh(\tau)\right] \, , \non\\
\txi_8 & = -(X_2 - P X_7)\Big[\coth(\tau) -\tau - 2\tau \csch^2(\tau)
  + \tau^2 \coth(\tau) \csch^2(\tau)\Big] \, .\non
\end{align}
}
The solution of~\cite{Kuperstein:2003yt} corresponds to the subset
$X_2 = P X_7$.\footnote{The relation with the notation of~\cite{Dymarsky:2011pm} is:
$ X_2^{here} = 2 X_2^{there}$, $X_7^{here} = - 2 X_7^{there}$.}
The modes $\tilde \phi^a$ are given by integral expressions
which can be evaluated numerically. In the following we will need the
UV behavior of one particular mode
$\tphi_5$, which corresponds to first-order perturbation of the PT scalar $f$ and which enters in the UV behavior of D5 charge~\eqref{QD5UV}. From~\cite{Bena:2011wh} we have
\begin{equation}
\tphi_5 = \tphi_5(\infty)+
\mathcal{O}(r^{-1}) \, ,
\end{equation} 
where (see for example eq.(93) of~\cite{Bena:2011wh} with $X_1=0$)
\begin{equation}\label{Y7UVkuso}
\tphi_5(\infty) \approx -19.5477\, P (2X_2 + PX_7) \, .
\end{equation}
We also need the expression for the mode $\tphi_2$ (perturbation of the metric mode $y$):
\begin{equation}
\tphi_2 = -\frac{32}{\sinh \tau} \int \frac{\sinh u \, \tilde 
  \xi_2(u)}{(\cosh u \sinh u-u)^{2/3}}\, du\, ,
\end{equation}
for which we could not find an analytic expression.
We now compute the relevant integrals that enter the  mass formula derived in the previous
section. First we would need the explicit expression for $\delta V_{\tilde z}$. By
plugging~\eqref{xi2gauginomasses} into~\eqref{Vzxiphi} we obtain
\begin{align}
\delta V_{\tilde z}& = -\frac{4\cosh \tau}{\sinh^2\tau} \tilde\phi_2 + \frac{8 P}{\sinh^4\tau(\cosh \tau \sinh\tau - \tau)^{2/3}}\Big[ 24\tau^2 -6 +4 \cosh(2\tau)+2\cosh(4\tau)\non\\
&\quad -26\tau \sinh(2\tau)+\tau\sinh(4\tau)\Big]X_7 -\frac{192
  (\cosh\tau\sinh\tau - \tau)^{4/3}}{\sinh^4\tau} X_2 \non\, .
\end{align}
Next, we evaluate the integral from \eqref{massfinalexprs} numerically
\begin{equation}\label{intdeltavgm}
\int_0^{\infty} \delta V_{\tilde z} \tilde z_0^2 d\tau\approx - 19.5477\, (2X_2+ P X_7) \, .
\end{equation}
At this point we recognize that the result is precisely the value of $ \tilde
\phi^5(\infty)/P$ found in~\eqref{Y7UVkuso}. Thus we have checked numerically that a small miracle happens and the contribution of $\delta V_{\tilde z}$ cancels the contribution from the new boundary conditions leaving the baryonic branch flat at the origin, in agreement with the field theory analysis 
\begin{equation}
 m^2_U= \frac{1}{ P\kappa_0 }(P \delta \tau + \tphi_5(\infty))= 0\ .
\end{equation}

In fact the field theory suggests that the baryonic branch will remain flat at linear order in $m_i$ everywhere along the branch, not only at the origin. That simply means that every solution from the BGMPZ family should admit two linearized perturbations which change leading UV boundary conditions only for relevant dimension $3$ operators (masses of gauginos). In other words linearized solutions found in \cite{Kuperstein:2003yt,Dymarsky:2011pm,Bena:2011wh} should exist for any value of $U$. It would be very interesting to develop a technique similar to \cite{Borokhov:2002fm} to systematically study linearized perturbations around any SUSY background  and find the linearized solutions dual to gaugino mass perturbations for any value of $U$ explicitly.

Another interesting question is what happens with the baryonic branch beyond the linear order in $m_i$.\footnote{This question was raised by I. Klebanov.} There is no reason to believe the branch will remain flat anymore as one would expect contributions of the sort
\bea
m^2_U\sim |m_i|^2+O(m_i^3)\ .
\eea
The sign of the emerging mass $m^2_U$ is not clear. If it is positive that would mean the locus of the baryonic branch remains as a stable vacuum in the theory. Otherwise the theory exhibits a run-away behavior. To understand in field theory  which scenario would take place is a difficult problem because of non-holomorphic nature of the  $|m_i^2|$ term. At the same time the answer can be obtained on the gravity side, by first constructing the background dual to the gaugino mass perturbation beyond linear order (this can be done either in perturbation theory in $m_i$ or numerically) and then using the machinery of section \ref{upliftmachinery}.

\section{Uplift of baryonic branch by (anti)D3-branes }\label{sec:susy}

We now turn to the uplifting by D3 or anti-D3 branes. We first perform the computation of the emergent mass in the closed string channel by applying the method described in the previous sections. We then look at the same mass from the open string perspective, and compare the results.

\subsection{Uplift from supergravity}
In this subsection we compute the mass $m^2_U$ (up to some universal coefficient), once D3 or anti-D3
branes are added to the KS background. 

We begin with the uplifting of the baryonic branch in the presence of
mobile D3-branes. While D3 branes are not BPS objects on the baryonic
branch, they become BPS at the origin $U=0$. This greatly simplifies the mass formula~\eqref{massfinalexprs}, since as was advertised in section \ref{bcsection},  $\delta V_{\tilde z}$ vanishes for a
regular supersymmetric perturbation. The emergent mass  is then
given simply by the shift in $\tau$ coordinate:
\begin{equation}
\delta \tau^{D3} = m^2_{D3} \kappa_0
\, .
\end{equation}
Using \eqref{taur} and \eqref{conjecture} the shift $\delta \tau$ can be easily calculated to be
\begin{equation}
  \delta \tau^{D3} = \frac{2\pi p}{M^2}  \, .
\end{equation}
Accordingly, the result for the mass
$ m^2_{D3}$ at the linear order in $p$ is 
\begin{equation}\label{D3inducedmass}
 m^2_{D3} = \frac{\delta \tau^{D3}}{\kappa_0}\approx  0.155011 \,
\frac{p}{M^2} \approx 0.00968816\, \frac{p}{P^2}\, .
\end{equation}


We now turn to the uplift by anti-D3 branes. The solution
corresponding to linearized backreaction of anti-D3 branes on the
Klebanov-Strassler geometry has been formally obtained in terms of
integrals in~\cite{Bena:2009xk,Dymarsky:2011pm} and was simplified and fully evaluated numerically
in~\cite{Bena:2011wh}. In what follows we will rely heavily on the numerical results of~\cite{Bena:2011wh}. More
generally, we will compute the mass for almost general non-supersymmetric
perturbation of the deformed conifold, preserving the $SU(2)\times
SU(2)\times \mathbb{Z}_2$ symmetries of the KS background. Such perturbations are parametrized by integration constants denoted $X_1, \dots X_8$. We then
impose the anti-D3 boundary conditions discussed
in~\cite{Dymarsky:2011pm,Bena:2011wh} to fix all the integration constants in terms of the number $p$ of anti-D3 branes and hence get the emergent mass in case of perturbation by the anti-D3-branes.
For simplicity in this section we put $P=1$.

We first need to evaluate the integral of $\delta V_{\tilde z}$ in~\eqref{massfinalexprs}. The numerical procedure is quite
involved but straightforward yielding 
\begin{equation}
\int_0^{\infty} \delta V_{\tilde z} \tilde z_0^2 d\tau \approx -48.6843  X_1 - 39.0955 X_2 + 22.6041 X_4 - 11.9448  X_6 - 
 19.5477 X_7 \, .
\end{equation}
We are setting $X_3 = 0$ since this integration constant corresponds to a highly divergent UV perturbation not induced by (anti)D3-branes presence.
Note that when $X_1=X_4=X_6=0$ we recover the value for the gaugino
mass solution~\eqref{intdeltavgm}.
By enforcing the anti-D3 boundary conditions described in eq. (74) of~\cite{Bena:2011wh}, we get the
following result
\begin{equation}
\int_0^{\infty} \delta V_{\tilde z} \tilde z_0^2 d\tau  \approx 2.98167\,  p\, . 
\end{equation}

We would also need the boundary term $\delta
\tau$, which is given by the value of the perturbation  mode $\tphi_5$ at infinity. The general expression is found by expanding in the UV the integral form of~\cite{Bena:2011wh}, which results in
\begin{equation}
\phi_5(\infty) \approx 64.9006 X_1 - 39.0955 X_2 + 16.525 X_4 + 0.213416 X_6 - 
 19.5477 X_7 - 36.4747 X_8 \, .
\end{equation}
For anti-D3-brane perturbation one finds:
\begin{equation}
\tphi_5(\infty)\approx 2.21552\, p \, .
\end{equation}
Combining all contributions together, we arrive at the final result for the anti-D3 branes:
\begin{align}
\delta \tau + \int_0^{\infty} \delta V_{\tilde z} \, \tilde z_0^2 d\tau \approx
0.766142 \, p \, .
\end{align}
Eventually we find 
\begin{equation}
m^2_{\overline{D}3} \approx  0.00446514 \, p\, .
\end{equation}

The masses $m^2_{D3}, m^2_{\overline{D3}}$ we found from supergravity are related to $m^2_U$ defined in field theory by a normalization coefficient which depends on the definition of $\hat{U}$. To get rid of this dependence we can consider the ratio
\begin{equation}\label{massratio}
\frac{m^2_{D3}}{m^2_{\overline{D}3}} \approx 0.512567 \, .
\end{equation}
In the next section, we will compute the same ratio using probe
approximation for D3 and anti-D3-branes.

\subsection{Probe computation of the induced masses}
Now we would like to repeat the calculation of mass $m^2_U$ in presence of D3 and anti-D3-branes in the open string channel. This calculation consists of two parts. First part is to calculate the potential along the baryonic  branch $V(U)$ emerging  due to the presence of the branes. At linear order in the number of branes this calculation is rather trivial and was done in \cite{DKS}. For $p$ D3-branes the potential is (see (15.2) of~\cite{DKS})
\begin{equation}
V_{D3}(U)=p {T_3\over \gamma}{U^2\over e^{-\Phi(0)}+1} \, ,
\end{equation}
while for $p$ anti-D3 it is  (see (14.10) of~\cite{DKS})
\begin{equation}
V_{\bar D3}(U)=p{T_3\over \gamma}{U^2\over e^{-\Phi(0)}-1}\, .
\end{equation}
Here $\Phi(0)$ is the $U$-dependent value of dilaton at the tip $\tau=0$.

Second part would be to calculate the kinetic term $K(U)$ associated with the motion along the baryonic branch (in other words $K(U)$ is a metric on the moduli space $ds^2=K(U) dU^2$). At the zeroth order in $p$ this calculation does not depend on $p$ and only depends on peculiarities of the BGMPZ solutions (or GHK mode for $K(0)$). Once $K$ is know the mass would be given by the usual $m^2_U=V''(0)/2 K(0)$.

A general method of calculating $K$ was put forward in \cite{Shiu:2008ry, Douglas:2008jx, Frey:2008xw}. It would be an interesting exercise to calculate $K(0)$ (and more generally $K(U)$) to match $m^2_{D3}, m^2_{\overline{D}3} $ found in the previous subsection. In this paper we prefer to avoid calculating $K$ by considering the ratio 
$
m^2_{D3}/ m^2_{{\overline{D}3}}=V_{D3}''(0)/V_{{\overline{D}3}}''(0)
$
which is $K$-independent.

Using the small $U$ expansion for dilaton  (see (13.14) of~\cite{DKS})
\begin{equation}
\phi(\tau=0,U)=-2^{-11/3}U^2 I(0)+\lambda U^4+{\mathcal O}(U^6)
\end{equation}
(here $I(0)\approx 0.718050$ is the value of the KS warp-factor at the tip
and $\lambda$ is some numerical constant), the desired mass ratio can be found to be
\begin{equation}
\frac{m^2_{\overline{D}3}}{m^2_{D3}}={2^{25/3}\lambda\over I(0)^2}-1\ .
\end{equation}
In order to calculate $\lambda$ we need to express the dilaton at the
tip in terms of $U$. A general formula expressing the dilaton through other fields is given by  (52) of~\cite{Dymarsky:2009fj}. Upon expanding near $\tau=0$ one finds
\begin{equation}
e^{4\Phi(0)}=-72{(1-2\xi)^3\over U^3}\ .
\end{equation}
Here  
\begin{equation}\label{IRxiU}
\xi(U) = \frac12 + \frac{U}{4\, 2^{2/3}} - \frac{I_0 U^3}{24\, 6^{2/3}} +
\frac{\mu}{2} U^5 + \mathcal{O}(U^7)\, 
\end{equation}
is an IR parameter that parametrizes solutions from the BGMPZ family (we apologize for possible clash of notations with $\tilde\xi_a$ from section 3).
Combining all together we arrive at
\begin{equation}\label{massratiomu}
\frac{m^2_{\overline{D}3}}{m^2_{D3}}=\mu {2^{1/3}3^{2/3}384\over I(0)^2}-{7\over 3}\ .
\end{equation}

The unknown constant $\mu$ defined
in~\eqref{IRxiU} is to be  calculated numerically.
A detailed account of this calculation can be found in the  appendix~\ref{appnumerics}, while here we
just state the final result
\begin{equation}
\mu \approx 0.00219499 \, .
\end{equation}
With this value the desired mass ratio turns out to be 
\begin{equation}
\left. \frac{m^2_{D3}}{m^2_{\overline{D}3}}\right|_{probe} =\left( \mu \frac{2^{1/3} 3^{2/3}
  384}{I(0)^2} - \frac73\right)^{-1}\approx 0.512567 \, .
\end{equation}

The result of the probe computation is in a full agreement
with the supergravity result~\eqref{massratio}. This confirms that our conjecture \eqref{conjecture} together with the interpretation of boundary conditions (\ref{alpha},~\ref{beta}) is correct. Furthermore this means the linearized mode describing backreaction of anti-D3-branes on the conifold (gravity background dual to the KPV state) passes this comprehensive test with the flying colors.

\section{Discussion}

In this paper we calculated the uplift of the baryonic branch (the mass of the scalar associated with the motion along the baryonic branch) in case the KS background is modified by some $\IZ_2$-even perturbation. Some of our results hold for perturbation of any magnitude, but we mostly focus on linear effects due to  infinitesimally small gaugino masses and presence of D3 or anti-D3-branes. In case of gaugino masses we found, in agreement with the field theory expectation, that the baryonic branch remains flat i.e.~the corresponding scalar is massless. An interesting question for the future would be to understand what happens with the baryonic branch beyond the linear order in gaugino masses. 

The rest of the paper deals with the baryonic branch uplift whenever D3 or anti-D3-branes are present. The latter case is very interesting because it provides a way to probe recently found linearized solution describing backreaction of anti-D3-branes placed at the tip of the deformed conifold. Our calculation shows that this solution passes an elaborate test and reproduces 
the emergent mass calculated using probe approximation precisely. This is a very encouraging confirmation that the boundary conditions and the resulting linearized solution formulated and found in \cite{Bena:2009xk, Dymarsky:2011pm, Bena:2011wh} is correct.

Now we would like to discuss what this means for the status of this solution, in particular if our finding  provides any new evidence proving existence of the meta-stable KPV state or reliability of the dS compactification scenario of String Theory. 

First of all we would like to emphasize that the original argument of \cite{KPV} concerns the regime $g_s p \ll 1$ i.e.~when the number of anti-D3-branes $p$ is small.
A more subtle question is if the meta-stability continues for $g_s p \gg 1$ while $p\ll M$ and if the corresponding state admits a dual supergravity description. A number of arguments based on the standard curvature estimate $ R^4\sim (g_s p)\alpha'^2$ and comparison with the related case of Polchinski and Strassler \cite{PS} suggest that there should be such a  supergravity background, but the necessary complexity of the corresponding solution (in particular angular dependence) has hampered all attempts to construct it so far. Moreover, in an effort to bypass this complexity, a search for a KPV-like polarization channel for the fully backreacted but smeared anti-D3 solution has been recently performed in~\cite{Bena:2012vz} albeit with  negative results.
Another issue regarding the smeared solution which has been recently investigated is the presence of singularities in three-form fluxes\footnote{See for example~\cite{Antibranepapers} for an account of the vivid debate on the interpretation of such singularities.}.

Unfortunately our analysis can not contribute in a definite way to the problem of existence of the self-consistent fully nonlinear holographic background dual to the KPV state when $g_s p \gg 1$ and $p\ll M$. The reason is the linear nature of the tested solution. This should be understood in the following way. Let us consider for example, a field theoretic configuration describing 
a  dense star. So far the star has not collapsed into a black hole the density of matter is finite everywhere and one expects to find a smooth classical configuration of fields (including gravity) describing it. There is no need to introduce any singularity, like the one in the center of the Schwarzschild solution,  which strictly speaking make sense only in quantum gravity. However at the linear level one oversees the  interaction (gravitational and all other) of the matter forming the star and hence know nothing about its spatial distribution. Looking from afar, all matter sits in the same point and the resulting  Newtonian (electrical etc.) potential $\varphi=G M_{\rm star}/r$ is clearly singular. But certainly this is not a sign of any inconsistency. Moreover linear solution correctly describes leading interaction between the star and other objects. At the same time correct linear solution obviously does not guarantee that the matter composing the star is capable of resisting the gravitational collapse and that the corresponding smooth field configuration exists. This can only be checked if one goes into full non-linear regime and either finds the corresponding smooth solution or shows that is not possible.  

The case of anti-D3-branes is very similar. In this paper we checked that the linear solution is indeed correct, but that is simply not enough to say anything about full non-linear regime. Moreover the case of backreacting anti-D3-branes is more complicated than the collapsing star (which is spherically symmetric)  because it is expected to break geometrical symmetries of the problem due to emerging special $\IS^2$ with NS5-brane wrapped around it. 



\vspace{0.5cm}
\noindent {\bf Acknowledgements}:
 \noindent 
We would like to thank C.~Ahn, I.~Klebanov, T.~Tesileanu, and B.~Safdi for collaboration on the early stages of this project and to M. Gra\~na, Z.~Komargodski, L.~McAllister, and N.~Seiberg for discussions. AD would like to thank International Institute of Physics (Federal University of Rio Grande do Norte) for hospitality while part of this work was done. AD gratefully acknowledges support from a Starting Grant of the European Research Council (ERC STG grant 279617) and the grant RFBR 12-01-00482. The work of SM is supported by a Contrat de Formation par la Recherche of CEA/Saclay, by the ERC Starting Independent Researcher Grant 259133 -- ObservableString and by the ERC Advanced Grant "Strings and Gravity" (Grant.No. 32004).

\appendix
\section{Derivation of $V_{\tilde z}$}\label{appVzder}

We derive the equation of motion~\eqref{zfulleom}
 for the mode $\tilde z$ by following the procedure of~\cite{Benna:2007mb}, namely by linearizing ten dimensional Einstein equations. Instead of giving full details of this derivation, we show here a quick way, mentioned in~\cite{GHK}, to derive the potential $V_{\tilde z}$ in~\eqref{vz} from the one-dimensional
effective action of Papadopoulos and
Tseytlin (PT)~\cite{PT}. The metric of the ansatz is
\begin{equation}
ds_{10}^2 = e^{2A+2p-x}ds_{1,3}^2 + e^{-6p-x}d\tau^2 + (e^{x+g}+a^2 e^{x-g})(e_1^2+e_2^2) +e^{x-g}(\tilde\epsilon_1^2+\tilde \epsilon_2^2)+e^{-6p-x}\tilde\epsilon_3^2 \, .
\end{equation}
This metric preserves the $\mathbb{Z}_2$ symmetry which interchanges the two $S^2$s only if $e^{g}+a^2 e^{-g} = e^{-g}$.
Since we are interested in an expansion around the $\mathbb{Z}_2$
symmetric solution we define
\begin{align}
e^{g} &= \frac{1}{\cosh y - c z} \, ,\\
a &= \frac{\sinh y}{\cosh y - c z} \, ,
\end{align}
where $c$ is a $\mathbb{Z}_2$-breaking parameter. Next, we derive an
effective action for $z$ at the $c^2$ order. We note that the field $\chi$
enters only through its derivative, so we eliminate it using the EOM
\begin{equation}
\chi ' = e^{-y} z ( f' + e^{2y} k') \, .
\end{equation}
The effective action then takes the form
\begin{align}
\mathcal{L}_z&=\frac{z^2 e^{4A-8p-4x-2y-\Phi}}{4(1+e^{2y})^2}\bigg[ 2
e^{2(6p+x+y+\Phi)}(e^{4y}-1)P F +e^{2(6p+x)}(e^{2y}+1) f'^2 \non \\
& \quad+
e^{2(6p+x+y)}(e^{2y}+1)^2 f' k' + e^{2(6p+x+2y)+2y}
(e^{2y}+1)k'^2\non \\
&\quad - e^{2y+\Phi}\Big(2(e^{2y}+1)(1+e^{2y}-2e^{6p+2x+y}+2
e^{12p+2x+2y+\Phi} P^2) + e^{4(3p+x)} (e^{2y}-1)^2 y'^2\Big) \bigg] \non \\
&\quad -\frac14 e^{4A+4p}z' \big(z' -2z y' \tanh y\big) \, .\label{zeffaction}
\end{align}
Now we define
\begin{equation}
\tilde z = 4 z e^{2A+2p} \, ,
\end{equation} 
so that the equation of motion for $\tilde z$ derived from~\eqref{zeffaction}
is
\begin{equation}
\tilde z'' - V_{\tilde z} \tilde z = 0 \, ,
\end{equation}
with $V_{\tilde z}$ given by~\eqref{vz}. 
 One can check that this equation agrees with the one obtained from Einstein equations when $m^2=0$.

\section{Particulars of numerical integration }\label{appnumerics}

In this section we describe the numerical procedure we used to compute the parameter $\mu$ in~\eqref{IRxiU}.
We use a shooting technique to numerically evaluate two particular modes of the BGMPZ family of solutions, $a(\tau)$ and $v(\tau) = e^{6p+2x}$. They are determined by a coupled system of ODEs~\cite{BGMPZ}:
\begin{align}
a' & = -\frac{a \sinh\tau (\tau + a
  \sinh\tau)}{\tau\cosh\tau-\sinh\tau}-\frac{1}{v}\Big[\sqrt{-1-a^2-2a\cosh\tau}(1+a\cosh
\tau)\csch \tau\Big]\, ,\label{ODEav}\\
v' & = -\frac{3a\sinh\tau}{\sqrt{-1-a^2-2a\cosh\tau}}\non \\
&\quad +
v \Big[-a^2\cosh^3\tau+2a \tau \coth\tau+a\cosh^2\tau
(2-4\tau\coth\tau) +\tau
\csch\tau \non\\
&\quad +\cosh\tau(1+2a^2-(2+a^2) \tau\coth\tau)\Big]/
\Big[(1+a^2+2a\cosh\tau)(\tau\cosh\tau-\sinh\tau)\Big]\, .\non
\end{align}
Since it is much easer to get perturbative series expansion in the IR rather than in the UV, we start by computing Taylor series for the functions $a(\tau)$ and $v(\tau)$ for small
$\tau$, up to a large number of terms, by solving the system~\eqref{ODEav}. The coefficients are polynomial functions of the parameter $\xi$:
\begin{equation}\label{Taylorav}
a(\tau,\xi) = -1 + \sum_{l=1}^{l^{\star}} a_l(\xi) \tau^{l}+\mathcal{O}(\tau^{l^{\star}+1}) \, , \quad
v(\tau,\xi) =  \sum_{l=1}^{l^{\star}} v_l(\xi) \tau^{l} +\mathcal{O}(\tau^{l^{\star}+1}) \, .
\end{equation}
We can easily compute the series for up to $l^{\star}=30$; the first terms
being
\begin{align}
a(\tau) &= -1 + \xi \tau^2
+\frac{1}{60}(-3+29\xi-114\xi^2+36\xi^3)\tau^4 + \mathcal{O}(\tau^6)
  \, , \label{IRseries}\\
v(\tau) &=  t + \frac{1}{120}(5-84\xi+84\xi^2) \tau^3
+\mathcal{O}(\tau^5) \, .\non
\end{align}
These series approximate the functions
$a(\tau)$ and $v(\tau)$ with an extremely high precision in the IR. 
Next, we choose a particular value of $\xi$ close to $\xi=1/2$ (that corresponds to $U=0$), which we call $\xi^{\star}$ and we calculate 
$a(\tau_{IR})$ and $v(\tau_{IR})$ for some particular $\tau_{IR}$ using~\eqref{Taylorav}.
The value of $\tau_{IR}$ is chosen such that $\tau_{IR}$ is sufficiently smaller than $1$ and series \eqref{Taylorav} converges rapidly. For example $\tau_{IR}=1/10$. The values of $a,v$
at $\tau_{IR}$ are used as boundary conditions to numerically evaluate with Mathematica \texttt{NDSolve} the
ODEs~\eqref{ODEav} from $\tau_{IR}$ to a point $\tau_{UV}$ in the moderate UV region. Lastly,  we need to determine the value of $U$ from the UV
series of $a(\tau)$. By solving~\eqref{ODEav} we get UV series up to sixth-order. The first terms are given below
\begin{align}
a(\tau) &= -2e^{-\tau} +U e^{-5\tau/3}(-1+\tau)  -\frac12 U^2
e^{-7\tau/3}(-1+\tau)^2 + \mathcal{O}(e^{-3\tau}) \, ,\label{UVseries}\\
v(\tau) & = \frac32 +\frac{9}{16}U^2 e^{-4\tau/3}(6-4\tau+\tau^2) +
e^{-2\tau}({\mathsf c}-6\tau) + \mathcal{O}(e^{-8\tau/3}) \, .\non
\end{align}
The constant ${\mathsf c}$ which first enters at order $e^{-2\tau}$ in the expansion of $v$ should be thought as a function of $U$.
 With these UV expansions we numerically find $U^{\star}$,
associated with $\xi^{\star}$, by matching the series with the result of \texttt{NDSolve} in $\tau_{UV}$. Finally using  
the expansion~\eqref{IRxiU} we determine $\mu$ though
\begin{equation}
\xi^{\star} = \frac12 + \frac{U^{\star}}{4\, 2^{2/3}} - \frac{I_0 (U^{\star})^3}{24\, 6^{2/3}} +
\frac{\mu}{2} (U^{\star})^5+\dots 
\end{equation}
Using different values of
$
\xi^{\star}= \left(\frac12 - 5\cdot 10^{-4} \, ; \frac12 -
  10^{-4} \right) $
and playing with the value of $\tau_{UV} = 15, 20$ and $\tau_{IR} = 1/10$ we find $\mu \approx 0.00219499(4)$, with the precision of about $10^{-6}$.


\providecommand{\href}[2]{#2}\begingroup\raggedright
\end{document}